\documentclass[cup5b]{caps}
\usepackage{graphicx}

\newcommand{\plotone}[1]{%
    \centering\includegraphics[width=\textwidth]{#1}}
\newcommand{\reffig}[1]{Fig.~\ref{#1}}
\newenvironment{references}%
    {\begin{list}{}{\setlength{\itemindent}{-1em}}}{\end{list}}

\begin{document}

\author[M. Cappellari et al.]{M. Cappellari,$^1$ R.C.E. van den Bosch,$^1$ E.K. Verolme,$^1$ R. Bacon,$^2$ M. Bureau,$^3$ \cr Y. Copin,$^4$ R.L. Davies,$^5$ E. Emsellem,$^2$ D. Krajnovic,$^1$ H. Kuntschner,$^6$ \cr R. McDermid,$^1$ B.W. Miller,$^7$  R.F. Peletier,$^8$ P.T. de Zeeuw$^1$\\
(1) Leiden Observatory (2) CRAL Observatory, Lyon (3) Columbia University, NY \\ (4) Institut de Physique Nucl\'eaire, Lyon (5) Oxford University\\ (6) European Southern Observatory (7) Gemini Observatory (8) Nottingham University}

\chapter{Dynamical Modeling of {\tt SAURON} Galaxies}

\section*{{\bf Abstract}}

We describe our program for the dynamical modeling of early-type galaxies observed with the panoramic integral-field spectrograph {\tt SAURON}. We are using Schwarz\-schild's numerical orbit superposition method to reproduce in detail all kinematical and photometric observables, and recover the intrinsic orbital structure of the galaxies. Since {\em catastrophes} are the most prominent features in the orbital observables, two-dimensional kinematical coverage is essential to constrain the dynamical models.

\section{Introduction}

We have observed a representative sample of 72 E, S0 and Sa galaxies (de Zeeuw et al.\ 2002) with the panoramic integral-field spectrograph (IFS) {\tt SAURON} (Bacon et al.\ 2001) mounted on the 4.2-m William Herschel telescope. The galaxies were observed to one effective radius,  with an effective spatial sampling of 0.8$''$. For these objects we extracted the stellar kinematics including the $h_3$ and $h_4$ Gauss--Hermite moments. Gas kinematics and line-strength distribution were measured as well.

We are in the process of constructing dynamical models for the E, S0 galaxies of the sample for which the kinematics and photometry are {\em consistent} with axisymmetry. Not to bias our conclusions the models need to be able to reproduce any general orbital distribution. For this we adopted Schwarzschild's (1979) numerical orbit-superposition method, which is able to fit all kinematical and photometric observations (Rix et al.\ 1997; van der Marel et al.\ 1998; Cretton et al.\ 1999). A similar approach was adopted by other groups to measure the black hole (BH) masses in galaxy nuclei (e.g.\ Gebhardt et al.\ 2003).

The implementation of Schwarzschild's method by van der Marel was adapted (Cappellari et al.\ 2002) for use with the Multi-Gaussian Expansion  parametrization of the galaxy surface brightness (Emsellem et al.\ 1994; Cappellari 2002) and was applied to IFS data (Verolme et al.\ 2002). There we showed that IFS observations are important to constrain model parameters such as the galaxy inclination, the stellar mass-to-light ratio and the BH mass. Triaxial Schwarzschild models have also been developed (see Verolme et al.\ 2003), and will be used to study the {\tt SAURON} kinematics. The goal is to derive intrinsic shapes, nuclear BH masses and orbital structure, to understand which models are preferred by the galaxies, and to set constraints on galaxy formation scenarios.

\section{Catastrophes in Orbital Observables}

Schwarzschild's dynamical modeling method consists of finding a positive linear combination of a representative set of orbital building-blocks so as to best fit the kinematical and photometric observables. Here we discuss the nature of orbital observables to better understand the results from the models.

In a three-dimensional time-independent potential, orbits which conserve three integrals of motion are called {\em regular} and can be reduced to translations on three-tori in a six-dimensional phase space (e.g.\ Binney \& Tremaine 1987). If there are no resonance conditions between the three frequencies of the motion on the torus, the trajectory of the system covers the torus surface with uniform density after an infinite amount of time. This property can be used to derive accurate time-averaged projected quantities for an orbit, without actually integrating the trajectory for an infinite time. It can be done from the knowledge of the transformation from the torus coordinates to the configuration space coordinates. Although the transformation is generally not known analytically, approximate methods exist for finding it from a short numerical integration of the orbit (e.g.\ Copin, Zhao, \& de Zeeuw 2000).

\begin{figure}
    \plotone{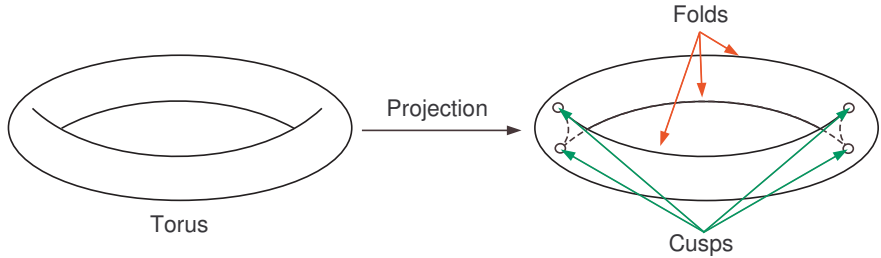}
    \caption{The most prominent features in the projection of a torus onto a plane are the singularities or catastrophes. Only the {\em fold} and {\em cusp} catastrophes can appear on the plane. This has applications in dynamics, as orbits lie on tori in phase-space.}
    \label{torus}
\end{figure}

At an even higher level, one can study the qualitative nature of orbital observables without even performing a single numerical integration of the orbit, using general results from {\em catastrophe} theory (e.g.\ Arnold 1992). When a manifold (e.g.\ a torus; \reffig{torus}) is projected onto a two-dimensional (2D) space, its projection is characterized by catastrophes. In 2D space only two stable catastrophes can appear, the {\em fold} and the {\em cusp}. The fold catastrophe is a curve on the plane, defined as the location where two of the inverse images in the projection merge and disappear. The cusp is a single point, lying at the intersection of folds, where three inverse images coalesce into a single one. If the density is constant on the manifold, the limiting surface density on the projection is $\Sigma(\mathbf{x})\propto |\mathbf{x}|^{-1/2}$ near the fold catastrophe and is $\Sigma(\mathbf{x})\propto |\mathbf{x}|^{-2/3}$  near the cusp. These singularities are integrable, as the mass is conserved by projection, and reduce to finite values in realistic situations (e.g.\ integrated over a pixel or convolved with a PSF), the density being higher at the cusps.

\begin{figure}
    \plotone{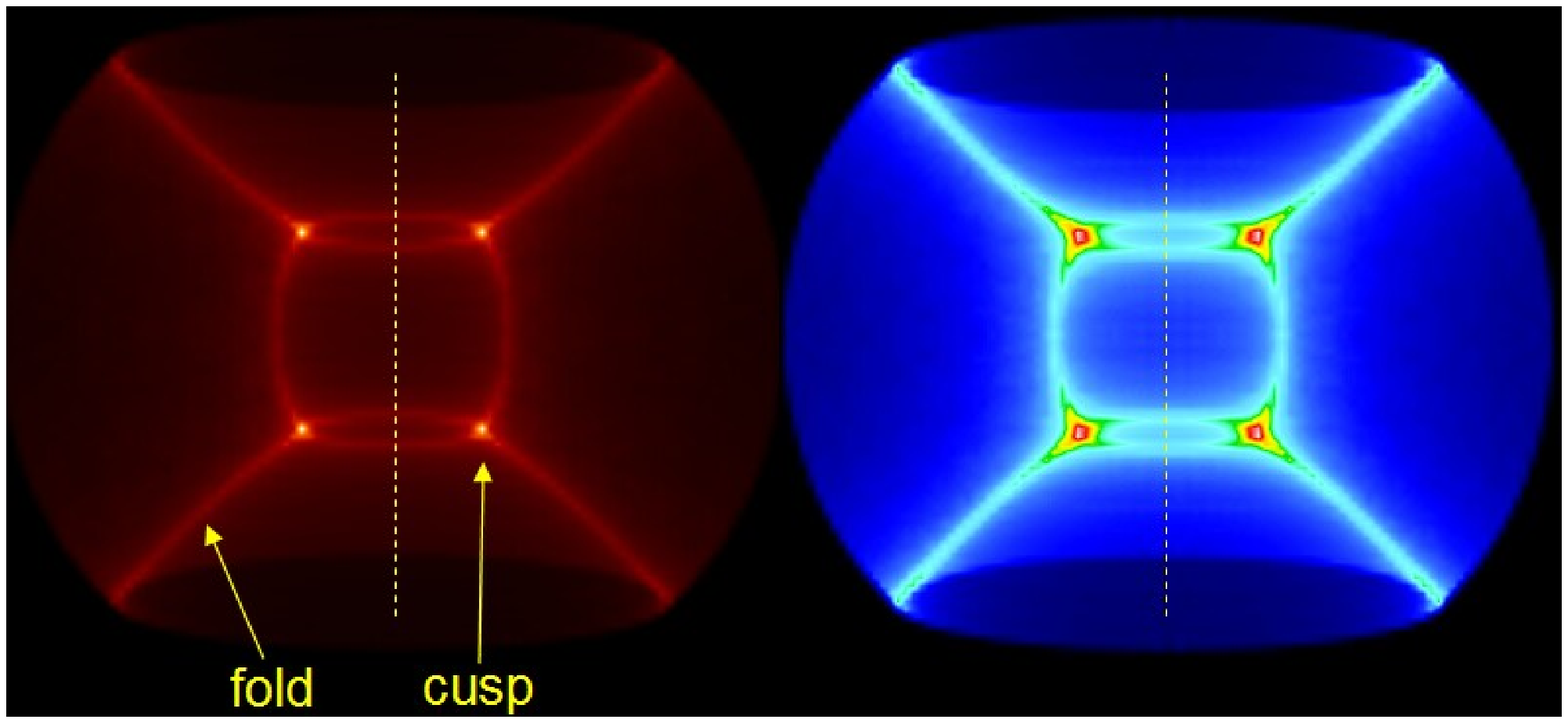}
    \plotone{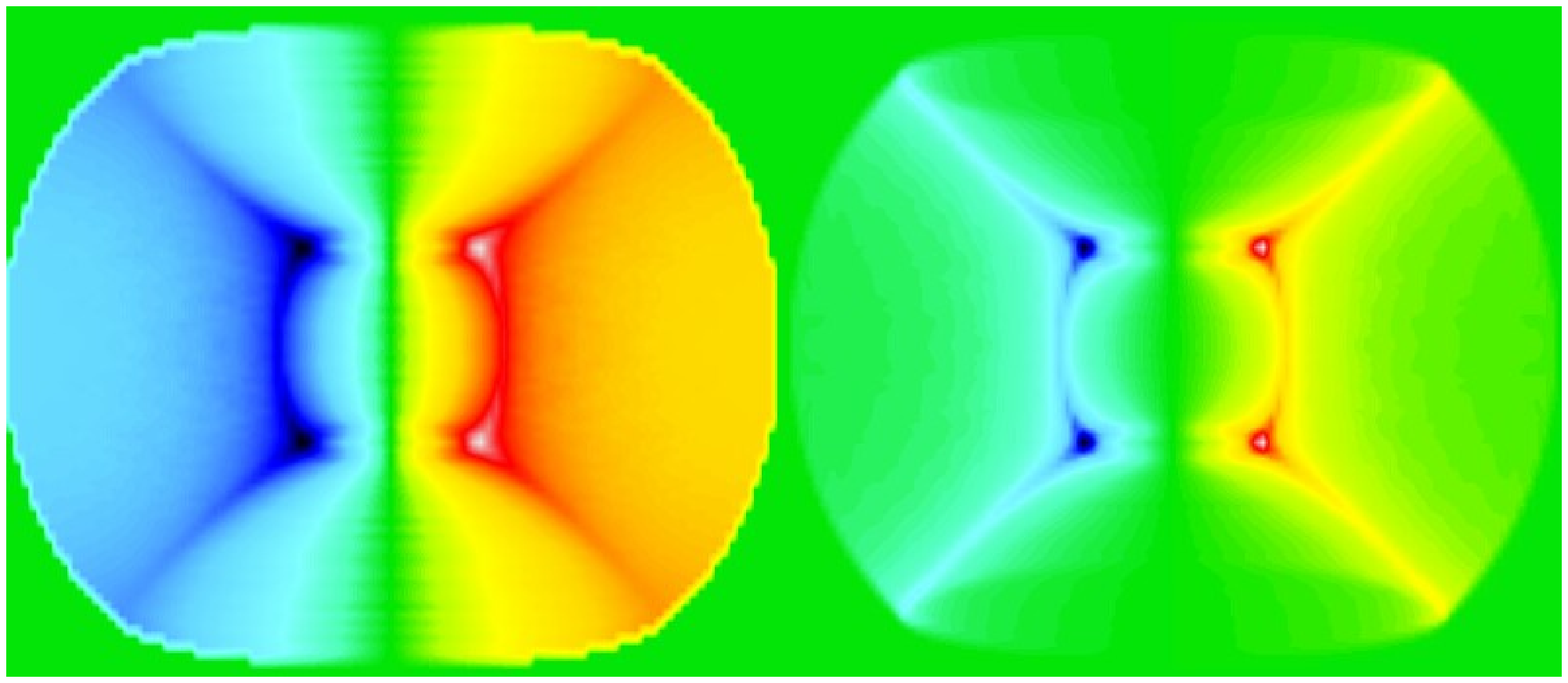}
    \caption{{\em Top Panels:} projected surface brightness of a regular non-resonant orbit in an axisymmetric potential. On the left a single-color linear colormap was used, while in all the other plots the same linear colormap of \reffig{ngc4473} was adopted. {\em Bottom Panels:} on the left the mean velocity of the above orbit is shown. On the right the velocity was weighted with the surface brightness: this is what enters into the computation of the observed mean velocity.}
    \label{axisymmetric}
\end{figure}

All the general features of the projection of a smooth surface on the plane can be applied to the projection of the uniform-density orbital torus. Fold and in particular cusp catastrophes will be the most prominent features in the projection of any regular orbit on the sky plane. This is precisely what one observes by numerically integrating a regular orbit in an axisymmetric potential, and computing orbital observables in a Monte Carlo fashion. \reffig{axisymmetric} shows the projected surface brightness and velocity field of a characteristic regular orbit. Its luminosity-weighted projected velocity is significantly different from zero only at the cusp positions, and similar behavior is observed for higher order moments. In an axisymmetric potential regular orbits are the dominant component, which means that catastrophes are important. One can draw the following conclusions, which we will discuss in more detail elsewhere:
\begin{itemize}
    \item A model composed mainly of regular orbits has a large flexibility in fitting general surface brightness distributions and complex kinematics. The model observables at a given position depend strongly only on the weights assigned to the orbits having a cusp at that position. Each orbit can be used to optimize the fit at the position of its cusps, and complex features (including noise in the data) can be reproduced.

    \item 2D kinematical coverage is essential to constrain the orbital structure in a galaxy from the observables. Orbital cusps can appear anywhere on the  projected image of the galaxy, and the weight of an orbit can be tightly constrained only when its cusp falls within an observed kinematical aperture;
\end{itemize}

\begin{figure}
    \plotone{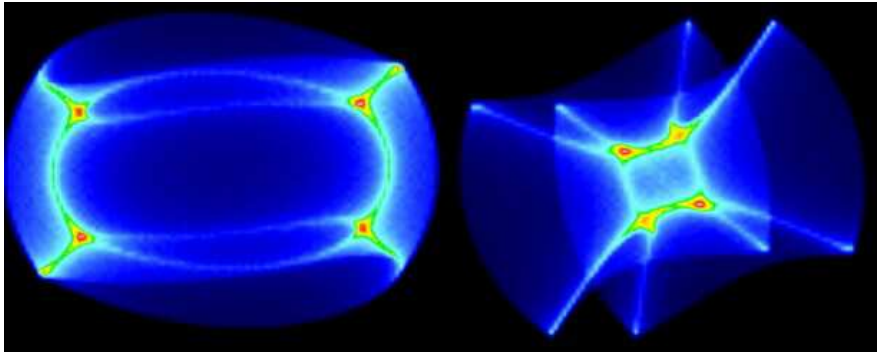}
    \caption{Projected surface brightness of a regular non-resonant tube orbit (left) and box orbit (right) in a triaxial potential.}
    \label{triaxial}
\end{figure}

In the case of triaxial galaxy potentials, regular non-resonant triaxial orbits will show the same general features as the axisymmetric case (\reffig{triaxial}).  Also here 2D kinematical coverage is crucial. However chaos can be significant in triaxial potentials with central singularities (Gerhard \& Binney 1985). Chaotic orbits do not lie on manifolds in phase space and their projections are {\em not} dominated by catastrophes.

\section{Dynamical Modeling of NGC~4473}

\begin{figure}
    \plotone{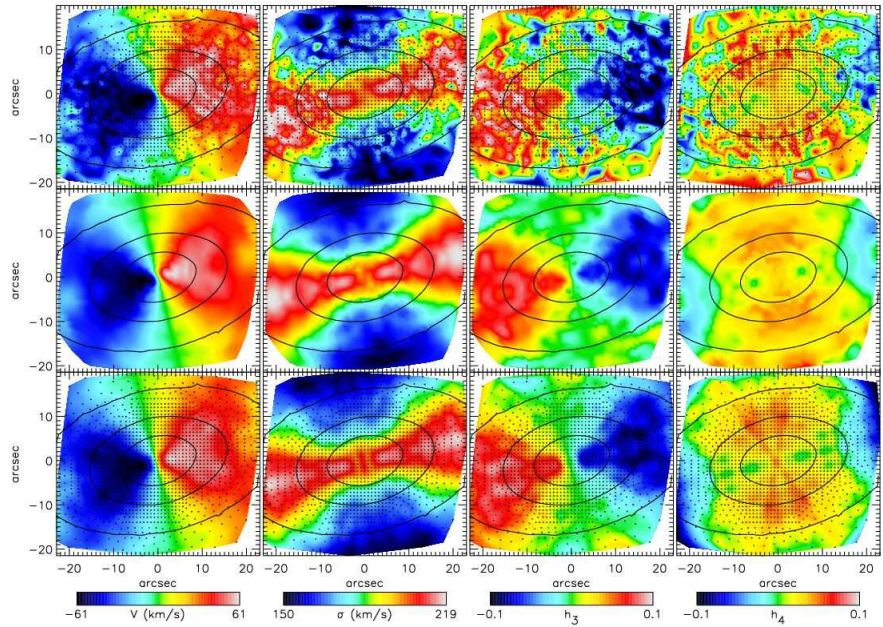}
    \caption{{\em Top Panels:} Voronoi 2D-binned and linearly interpolated {\tt SAURON} kinematics of NGC~4473. The bin centroids are indicated by the dots. Some representative galaxy isophotes are also shown. {\em Middle Panels:} to help the visual comparison of the model with the data, we show here a bisymmetric 16-terms Fourier expansion (e.g.\ Copin et al.\ 2002) of the data in the top panels. No axisymmetric model can fit the data better than this. {\em Bottom Panels:} as in the top panel for the best-fitting regularized axisymmetric dynamical model.}
    \label{ngc4473}
\end{figure}

As an example of the models we are constructing, we show in \reffig{ngc4473} a data-model comparison for the best-fitting axisymmetric Schwarzschild dynamical model to the {\tt SAURON} observations of the E5 galaxy NGC~4473. This galaxy was chosen because of its peculiar velocity dispersion field, remaining essentially constant along the galaxy major axis. For an optimal extraction of the kinematics the {\tt SAURON} datacube was first binned spatially to a constant S/N using the Voronoi 2D-binning method by Cappellari \& Copin (2003). Our Schwarzschild modeling software is able to take the irregular shape of the bins precisely and efficiently into account.

As expected from the above considerations, the model is able to reproduce the mean velocity $V$, the velocity dispersion $\sigma$, and the Gauss--Hermite moments $h_3$ and $h_4$, well within the errors, over the whole observed field. From the analysis of the model orbital structure we find that two components, of different angular momentum, are needed to fit the observations and explain the observed peculiar kinematics.

The analysis of the orbital structure of the galaxies in the sample we are modeling, will provide, as a function of morphological type and luminosity, information on the intrinsic shapes, the mass distribution and orbital structure, including decoupled components. The knowledge of the DF, in connection with the stellar population analysis we are performing, provides clues to the galaxy formation processes and the link between the formation of the galaxy and the formation of its central BH.

\end{document}